\documentclass[conference,letterpaper,10pt]{IEEEtran}

\usepackage[top=0.75in,bottom=1.05in,left=0.625in,right=0.63in]{geometry}
\usepackage[cmex10]{amsmath}
\usepackage{amssymb}

\usepackage{rotating}
\usepackage{epstopdf}
\usepackage{cite}
\usepackage[utf8]{inputenc} 
\usepackage[T1]{fontenc}
\usepackage{url}
\usepackage{titlesec}
\titlespacing*{\subsection}{0pt}{0pt}{0pt}
 \usepackage{mathtools}
\usepackage{ifthen}
\usepackage[switch]{lineno}
\usepackage{cite}
\usepackage[cmex10]{amsmath} 
\usepackage{amssymb}
\usepackage{float}

\usepackage{graphicx}
\usepackage{helvet}
\usepackage{titlesec}
\usepackage{titlesec}
\titlespacing*{\subsection}{0pt}{*0.5}{*0.5}   

\usepackage{titlesec}
\titlespacing*{\section}{0pt}{*1}{*0.7} 

\usepackage{epstopdf}
\usepackage{epsfig}
\usepackage{titlesec}
\usepackage{multirow}
\usepackage{amsthm}
\usepackage{setspace}

\usepackage{comment}
\usepackage{caption}

\usepackage{breqn}
\usepackage{graphicx}
\usepackage{stfloats,adjustbox}
\usepackage{url}
\usepackage{xcolor}
\usepackage{multirow}
\usepackage{bm}
\usepackage{bbm}
\usepackage{amsmath}
\usepackage{enumitem}
\usepackage{subcaption}
\usepackage[english]{babel}
\usepackage{ragged2e}
\usepackage{blindtext}
\usepackage{caption, multirow, makecell}

\usepackage{fixltx2e}
\usepackage{longtable,eqnarray,makecell}
\usepackage{amsmath,amssymb}
\usepackage{threeparttable,stackengine}
\usepackage{tikz}
\usepackage{amsthm}
\newtheorem{example}{Example}
\usepackage{booktabs}
\newtheorem{thm}{Theorem}
\newtheorem{lem}{Lemma}

\newtheorem{defn}{Definition}

\newcounter{myitemcounter}
\begin{document}
\title{Function Correcting Codes for Maximally-Unbalanced Boolean Functions} 

\author{
Rajlaxmi Pandey\textsuperscript{1}, Shiven Bajpai\textsuperscript{2}, Anjana A Mahesh\textsuperscript{3}, B. Sundar Rajan\textsuperscript{1} \\
\textsuperscript{1}Department of Electrical Communication Engineering, Indian Institute of Science, Bengaluru, India \\
\textsuperscript{2}Department of Artificial Intelligence, Indian Institute of Technology, Hyderabad, India \\
\textsuperscript{3}Department of Electrical Engineering, Indian Institute of Technology, Hyderabad, India \\
\textsuperscript{1}\texttt{\{rajlaxmip, bsrajan\}@iisc.ac.in}, 
\textsuperscript{2}\texttt{ai24btech11030@iith.ac.in}, 
\textsuperscript{3}\texttt{anjana.am@ee.iith.ac.in}
}


\maketitle


\begin{abstract}
Function-Correcting Codes (FCCs) enable reliable computation of a function of a $k$-bit message over noisy channels without requiring full message recovery. In this work, we study optimal single-error correcting FCCs (SEFCCs) for maximally-unbalanced Boolean functions, where $k$ denotes the message length and $t$ denotes the error-correction capability. We analyze the structure of optimal SEFCC constructions through their associated codeword distance matrices and identify distinct FCC classes based on this structure. We then examine the impact of these structural differences on error performance by evaluating representative FCCs over the additive white Gaussian noise (AWGN) channel using both soft-decision and hard-decision decoding. The results show that FCCs with different distance-matrix structures can exhibit markedly different Data BER and function error behavior, and that the influence of code structure depends strongly on the decoding strategy.
\end{abstract}



\IEEEpeerreviewmaketitle
\section{Introduction}
\label{intro}

Error-correcting codes (ECCs) are traditionally designed to enable reliable recovery of the entire transmitted message over a noisy channel, under the assumption that every symbol of the message holds equal significance. However, in many settings, the receiver may be interested only in computing a specific function of the message, rather than decoding it entirely. In such cases, if the encoder has prior knowledge of the function of interest, it is possible to design codes that prioritize the accurate recovery of the function outcome, even in the presence of errors. This perspective motivates the study of \emph{function-correcting codes} (FCCs), which were introduced in \cite{lenz2023function}. For a function \( f \) and a positive integer \( t \), a systematic encoding is called an \((f,t)\)-FCC if it can protect the function value against at most \( t \) errors.

The authors in \cite{lenz2023function} laid the foundation for function-correcting codes (FCCs) by introducing them as irregular-distance codes, i.e., codes that satisfy a prescribed pairwise distance requirement between codewords. They focused on deriving upper and lower bounds on the optimal redundancy of FCCs for specific families of functions, such as locally binary functions, Hamming weight functions, and Hamming weight distribution functions. In addition, they presented optimal constructions of FCCs for several of these function classes. Their analysis was conducted under the assumption of a binary symmetric channel.

In~\cite{xia2024function}, the authors extended function-correcting codes to symbol-pair read channels by introducing function-correcting symbol-pair codes (FCSPCs). 
They analyzed the optimal redundancy of FCSPCs and established upper and lower bounds by relating them to irregular-pair-distance codes. 
Results were provided for both generic and specific functions.

More recently, Premlal and Rajan~\cite{premlal2025function} further developed the function-dependent graph framework introduced in~\cite{lenz2023function}, deriving tight bounds on the redundancy required for various function classes, including linear functions and bijections. They also highlighted connections to classical single-error-correcting codes and identified cases where coset-wise coding reduced to a lower-dimensional error correction problem. Ge \emph{et~al.}~\cite{ge2025optimal} further advanced function-correcting codes by analyzing the optimal redundancy for Hamming weight and Hamming weight distribution functions. Their work improved previous bounds and provided explicit constructions via a connection with Gray codes, achieving near-optimal redundancy for Hamming weight and identifying regimes with varying redundancy for the Hamming weight distribution function.

Singh \emph{et~al.}~\cite{singh2025b} further extended function-correcting codes to $b$-symbol read channels over finite fields. They introduced irregular $b$-symbol distance codes and provided bounds on optimal redundancy, along with a graphical representation of the construction problem. Their results showed that function-correcting $b$-symbol codes could achieve lower redundancy than classical $b$-symbol codes while maintaining reliability, offering insight into FCC behavior in more complex channel settings.  
Rajput \emph{et~al.}~\cite{rajput2025function} studied function-correcting codes for locally $(\rho,\lambda)$-bounded functions, provided upper bounds on redundancy, and gave conditions for optimality, applicable to general functions including Hamming weight distribution functions. Sampath \emph{et~al.}~\cite{sampath2025note} studied function-correcting codes for linear functions over $b$-symbol read channels. The work provided a Plotkin-like bound for FCCs in this setting, which reduced to the corresponding bound for symbol-pair channels when $b=2$. For linear bijective functions, this bound further reduced to the classical Plotkin bound for error-correcting codes, highlighting the connection between FCCs and traditional ECCs.

Liu and Liu~\cite{liu2025homogeneous} introduced function-correcting codes with homogeneous distance (FCCHDs), extending FCCs with Hamming distance to codes over finite rings. They established connections between the optimal redundancy of FCCHDs and the lengths of certain codes, derived bounds on redundancy for specific functions, and constructed FCCHDs for homogeneous weight functions and distributions, achieving optimal or near-optimal redundancy in several cases.
Ly and Soljanin \cite{ly2025} derived upper and lower bounds on the redundancy of function-correcting codes over finite fields and proposed an encoding scheme achieving optimal redundancy for sufficiently large fields.
Subsequently, Rajput et al. \cite{rajput2025data} extended the FCC framework by incorporating explicit data protection alongside function protection, proposing general constructions and bounds while showing that data protection can be added without increasing redundancy for several function classes.

\subsection{Notations}

We use \( \mathbb{N} \) to denote the set of all positive integers, and \( \mathbb{N}_0 \) to denote the set of non-negative integers. For any positive integer \( N \), the shorthand \( [N] \) refers to the set \( \{1, 2, \dots, N\} \). The set of real and complex numbers are denoted by \( \mathbb{R} \) and \( \mathbb{C} \), respectively.
\begin{figure}
 \begin{center}
 \captionsetup{justification = centering}
\includegraphics[width=0.75\columnwidth]{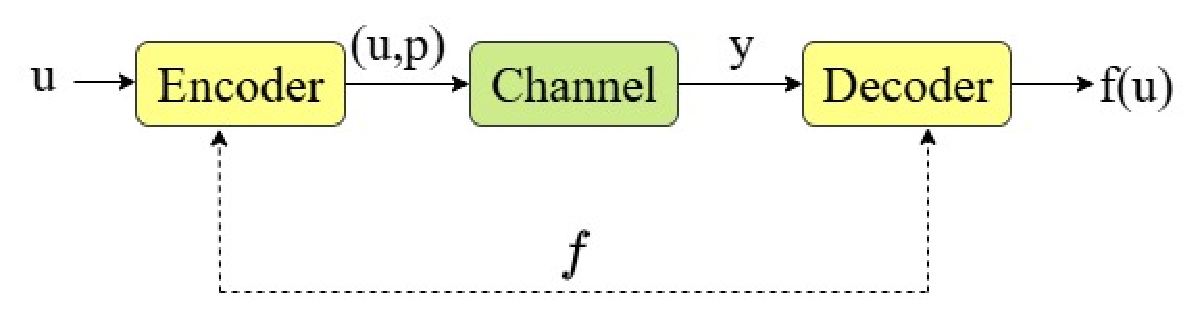}
 
    \caption{System model for the function-correcting code under noisy channel.}
    \label{fig:model}
    \end{center}
\end{figure}
The finite field with \( q \) elements, where \( q \) is a prime or a prime power, is denoted by \( \mathbb{F}_q \). In particular, \( \mathbb{F}_2 \) denotes the binary field. Vectors of length \( n \) over \( \mathbb{F}_q \) form the space \( \mathbb{F}_q^n \). A linear block code with length \( n \), dimension \( k \), and minimum Hamming distance \( d \) over \( \mathbb{F}_q \) is denoted as a \( [n, k, d]_q \) code.

We denote the Hamming distance by $d_H(\cdot,\cdot)$. The Hamming weight of a vector $\mathbf{x} \in \mathbb{F}_q^n$, denoted by $w_H(\mathbf{x})$, is the number of nonzero positions in $\mathbf{x}$. For any vector $\mathbf{u}_i$, the notation $(\mathbf{u}_i)^t$ denotes the vector obtained by repeating $\mathbf{u}_i$ consecutively $t$ times. For example, if $\mathbf{u}_i = 10$, then $(\mathbf{u}_i)^3 = 101010$.

A general (not necessarily linear) code of length \( n \), containing \( M \) codewords, and with minimum distance \( d \), is denoted as an \( (n, M, d) \) code. The function \( N(M, d) \) denotes the smallest length \( n \) for which such a code exists.

The notation $\mathbf{0}_k$ is used to denote the $k$-length all-zero vector. Similarly, the $k$-length all-one vector is denoted as $\mathbf{1}_k$. For a $k$-input Boolean function $f$, the subset of all $k$-length vectors for which $f$ evaluates to 0 and 1 are denoted as  $f^{-1}(0)$ and $f^{-1}(1)$, respectively, i.e., $f^{-1}(0)$ and $f^{-1}(1)$ are the pre-images of $0$ and $1$, respectively, under  $f$. 

\section{Preliminaries}
\begin{defn}
Let $f : \mathbb{F}_2^k \to S$ be a function, where $S$ denotes the image of $f$.
A systematic encoder $\mathsf{Enc} : \mathbb{F}_2^k \to \mathbb{F}_2^{k+r}$,
defined as $\mathsf{Enc}(\mathbf{u}) = (\mathbf{u}, \mathbf{p}(\mathbf{u}))$,
is called an $(f,t)$-function-correcting code (FCC) if, for all
$\mathbf{u}_1, \mathbf{u}_2 \in \mathbb{F}_2^k$ such that
$f(\mathbf{u}_1) \neq f(\mathbf{u}_2)$,
\[
d_H\!\left(\mathsf{Enc}(\mathbf{u}_1), \mathsf{Enc}(\mathbf{u}_2)\right) \ge 2t+1,
\]
where $d_H(\cdot,\cdot)$ denotes the Hamming distance.
\end{defn}
For a bijective function $f$, any function-correcting code reduces to a standard error-correcting code. 
For a constant function $f$, the trivial encoding $\mathsf{Enc}(\mathbf{u}) = \mathbf{u}$ achieves zero redundancy. 
Finally, without knowledge of $f$, only standard error-correcting codes can be applied.

\begin{defn}
Let $\mathbf{u}_1, \dots, \mathbf{u}_M \in \mathbb{F}_2^k$. 
For a function $f : \mathbb{F}_2^k \to \mathcal{S}$, the \emph{distance requirement matrix (DRM)} 
$D_f(t; \mathbf{u}_1, \dots, \mathbf{u}_M) \in \mathbb{Z}^{M \times M}$ is defined entry-wise as
\[
[D_f(t; \mathbf{u}_1, \dots, \mathbf{u}_M)]_{i,j} =
\begin{cases}
\left[ 2t + 1 - d_H(\mathbf{u}_i, \mathbf{u}_j) \right]^+, \\ \quad \text{if } f(\mathbf{u}_i) \neq f(\mathbf{u}_j),\\
0, \quad \text{otherwise},
\end{cases}
\]
where $d_H(\cdot, \cdot)$ denotes the Hamming distance and $[x]^+ = \max\{x,0\}$.
\end{defn}

\subsection*{Decoding Models}
We evaluate the performance of various FCCs under two standard decoding paradigms: hard-decision decoding and soft-decision decoding. In both cases, transmission occurs over an additive white Gaussian noise (AWGN) channel with binary phase-shift keying (BPSK) modulation, where bit 0 is mapped to \( +1 \) and bit 1 to \( -1 \).

In hard-decision decoding, the received noisy signal is first quantized to a binary vector by thresholding at zero, and decoding is performed by selecting the codeword from a predefined table that minimizes the Hamming distance to this binary vector.

In contrast, soft-decision decoding directly uses the real-valued output of the AWGN channel without quantization. Decoding is performed by selecting the codeword whose BPSK-modulated version is closest in Euclidean distance to the received signal.

Together, these approaches allow us to assess how FCCs perform for binary-valued functions when using different decoding strategies.

In this section, we first define maximally-unbalanced Boolean functions, an example of which is the multi-input OR function, and then characterize the number of distinct function correcting codes (FCCs) for maximally-unbalanced Boolean functions, and analyze the distance properties of these distinct FCCs. Towards characterizing distinct FCCs, we define the \emph{codeword distance matrix (Codeword DM)} of an FCC.
\begin{defn}[Maximally-Unbalanced Boolean Function]
    A Boolean function $f: \mathbb{F}_2^k \rightarrow \mathbb{F}_2$ is said to be maximally-unbalanced if either the pre-image of $0$ under $f$ is a singleton set or the pre-image of $1$ under $f$ is a singleton set, i.e., if $|f^{-1}(0)| = 1$ or $|f^{-1}(1)| = 1$.
\end{defn}

An example of the maximally-unbalanced Boolean function (MUBF) is the multi-input OR function defined below. 

\textit{Multi-input OR Function:}
Let $\mathbf{u} = (u_1, u_2, \dots, u_k) \in \{0,1\}^k$ be a length-$k$ binary vector. 
The \emph{$k$-input OR function} is defined as
\[
\vee_k(\mathbf{u}) = u_1 \lor u_2 \lor \cdots \lor u_k = \bigvee_{i=1}^{k} u_i,
\]
where $\lor$ denotes the logical OR.

In this paper, we consider only FCCs with systematic encodings, i.e., the encoding of any vector $\mathbf{u}_i \in \mathbb{F}_2^k$ is the $(k+r)$-length binary vector obtained by the concatenation of $\mathbf{u}_i$ and the parity vector $\mathbf{p}(\mathbf{u}_i) \in \mathbb{F}_2^r$ assigned to $\mathbf{u}_i$, i.e., $enc(\mathbf{u}_i)=[\mathbf{u}_i, \mathbf{p}(\mathbf{u}_i)]$.

For a maximally-unbalanced Boolean function $f: \mathbb{F}_2^k \rightarrow \mathbb{F}_2$, without loss of generality, assume that $|f^{-1}(0)| = 1$ and $|f^{-1}(1)| = 2^k-1$. Further, let $f^{-1}(0) = \{\mathbf{u}_0\}$. A valid $t$-error correcting FCC for this function $f$ needs to satisfy the condition 

\begin{equation}
\label{eq:MUBF_distReq1}
    d_H(enc(\mathbf{u}_0),enc(\mathbf{u}_i)) \geq 2t+1, \quad \forall u_i \in \mathbb{F}_2^k \setminus \{\mathbf{u}_0\}
\end{equation}. 

From \cite{lenz2023function}, we know that the optimal redundancy for a $t$-error correcting Boolean function is $r^* = 2t$. Specializing \eqref{eq:MUBF_distReq1} to optimal systematic $t$-error correcting FCCs for $f$ results in the following distance requirement. 
\begin{equation}
\label{eq:MUBF_distReq2}
    d_H(\mathbf{p}(\mathbf{u}_0),\mathbf{p}(\mathbf{u}_i)) \geq 2t+1 - d_H(\mathbf{u}_0,\mathbf{u}_i), \ \forall u_i \in \mathbb{F}_2^k \setminus \{\mathbf{u}_0\}.
\end{equation} 

Further, if we assume that $f^{-1}(0) = \{\mathbf{0}_k\}$, then the equation \eqref{eq:MUBF_distReq2} further simplifies to 
\begin{equation}
\label{eq:MUBF_distReq3}
    d_H(\mathbf{p}(\mathbf{0}_k),\mathbf{p}(\mathbf{u}_i)) \geq 2t+1 - w_H(\mathbf{u}_i), \quad \forall u_i \in \mathbb{F}_2^k \setminus \{\mathbf{u}_0\}
\end{equation}.

\begin{defn}[Codeword Distance Matrix (Codeword DM)]
Let \( \mathcal{C} = \{ \mathbf{c}_1, \mathbf{c}_2, \dots, \mathbf{c}_M \} \subseteq \mathbb{F}_2^n \) be a binary block code with \( M \) codewords of length \( n \).
The \emph{codeword distance matrix} of \( \mathcal{C} \) is the \( M \times M \) matrix \( H \) defined by
\[
H(i,j) = d_H(\mathbf{c}_i, \mathbf{c}_j),
\]
where \( d_H(\mathbf{c}_i, \mathbf{c}_j) \) denotes the Hamming distance between \( \mathbf{c}_i \) and \( \mathbf{c}_j \).
The matrix \( H \) is symmetric and satisfies \( H(i,i) = 0 \) for all \( i \).
\end{defn}

Two codeword distance matrices $H_1$ and $H_2$ are said to be \textit{distinct} if they differ in at least one entry. Since, the codewords DMs are symmetric matrices, the distinct codeword DMS will differ in an even number of entries, at least two.

\begin{defn}[Comparable Codeword Distance Matrices]

Two distinct codeword DMs $H_0$ and $H_1$ of size $M \times M$ are said to be comparable if $H_0(i,j) \leq H_1(i,j)$ or vice-versa for all $i,j \in \{1,2,\cdots M\}$. Further, if $H_0(i,j) \leq H_1(i,j)$, $\forall i,j \in \{1,2,\cdots M\}$, it denoted as $H_0 \leq H_1$.
\end{defn}

Two codeword DMs that are not comparable are said to be incomparable. 

\begin{defn}[Comparable Chain]
A set of $M \times M$ distinct codeword DMs, $H_1, H_2, \cdots, H_N$ are said to form a comparable chain if there exists an ordering $\{i_1,i_2,\cdots,i_N\}$of the $N$ codeword DMs such that $H_{i_1} \leq H_{i_2} \leq \cdots \leq H_{i_N}$ and the matrix $H_{i_N}$ is called the chain leader. 
\end{defn}

In the rest of this paper, we investigate optimal single-error correcting function correcting codes (SEFCCs) for maximally-unbalanced Boolean functions (MUBFs). Without loss of generality, we make the assumption that for the given MUBF $f: \mathbb{F}_2^k \rightarrow \mathbb{F}_2$, $|f^{-1}(0)| = 1$ and that $f^{-1}(0)$ is the $k$-length all-zero vector $\mathbf{0}_k$. 

\section{Main Results}

\begin{thm}
\label{Thm:NumFCC}
For a maximally-unbalanced Boolean function $f: \mathbb{F}_2^k \rightarrow \mathbb{F}_2$, the number of optimal FCCs with single-error correction capability is given by $$\mathcal{N}(f,t=1) =3^{\binom{k}{2}}\times4^{\big(2^k - \binom{k}{2} - \binom{k}{1}\big)}$$
\end{thm}

\begin{IEEEproof}
   Let $\mathbf{p}(\mathbf{u}_i)$ denote the parity vector assigned to a message vector $\mathbf{u}_i \in \mathbb{F}_2^k$. Since we consider optimal single-error-correcting FCCs for $f$, the parity vectors have length $2t=2$, i.e., $\mathbf{p}(\mathbf{u}_i) \in \mathbb{F}_2^2$.

Without loss of generality, assume that $|f^{-1}(0)| = 1$ and that $f^{-1}(0)$ corresponds to the $k$-length all-zero vector $\mathbf{0}_k$. For the vector $\mathbf{0}_k$, there are $2^2 = 4$ possible choices for the parity vector, namely $00$, $01$, $10$, and $11$.

Let $\mathbf{U}_i$ denote the set of all vectors in $\mathbb{F}_2^k$ with Hamming weight $i$, i.e.,
\[
\mathbf{U}_i \triangleq \left\{ \mathbf{u}_j \in \mathbb{F}_2^k \;\middle|\; w_H(\mathbf{u}_j) = i \right\}.
\]

Since $f$ is a maximally-unbalanced Boolean function and, under our assumption, $f^{-1}(0) = \{\mathbf{0}_k\}$, all remaining vectors in $\mathbb{F}_2^k$ belong to $f^{-1}(1)$. For a valid FCC, their encodings must satisfy the distance constraint in \eqref{eq:MUBF_distReq3}.

In particular, all vectors of Hamming weight $1$ must satisfy
\[
d_H\!\left(\mathbf{p}(\mathbf{u}_i), \mathbf{p}(\mathbf{0}_k)\right) \geq 2,
\quad \forall\, \mathbf{u}_i \in \mathbf{U}_1.
\]
Thus, for a given choice of $\mathbf{p}(\mathbf{0}_k) \in \mathbb{F}_2^2$, the parity vectors of all vectors in $\mathbf{U}_1$ are uniquely determined as $\mathbf{p}(\mathbf{0}_k)^c$.

Similarly, the parity vector for each of the $\binom{k}{2}$ vectors in $\mathbf{U}_2$ must be at a Hamming distance of at least one from $\mathbf{p}(\mathbf{0}_k)$ and can therefore be chosen from any of the three length-$2$ binary vectors distinct from $\mathbf{p}(\mathbf{0}_k)$.

Finally, for the remaining $2^k - 1 - \binom{k}{2} - \binom{k}{1}$ vectors, i.e., those in $\mathbb{F}_2^k \setminus (\mathbf{U}_1 \cup \mathbf{U}_2)$, the message vectors are already at a Hamming distance of at least $3$ from $\mathbf{0}_k$. Hence, their parity vectors may be chosen arbitrarily from $\mathbb{F}_2^2$.

Therefore, the total number of possible FCCs for $f$ with single-error-correction capability is
\[
\mathcal{N}(f,t=1)
= 4 \times 3^{\binom{k}{2}} \times 4^{\,2^k - 1 - \binom{k}{2} - \binom{k}{1}} .
\]

\end{IEEEproof}

\begin{lem}
\label{Lem:DistinctCDM}
    For a given MUBF $f: \mathbb{F}_2^k \rightarrow \mathbb{F}_2$, all the distinct codeword DMs corresponding to optimal single-error correcting FCCs can be achieved using a single choice of parity vector for the all-zero vector $\mathbf{0}_k$.
\end{lem}

\begin{IEEEproof}
    Consider an optimal SEFCC $\mathcal{C}_0$ for the given MUBF $f$ with the corresponding parity assignment $\mathbf{p}_{\mathcal{C}_0}(\mathbf{u}_i)$ denoted as $\mathbf{p}^{\mathcal{C}_0}(\mathbf{u}_i) = [p^{\mathcal{C}_0}_1(\mathbf{u}_i), \ p^{\mathcal{C}_0}_2(\mathbf{u}_i)], \forall \mathbf{u}_i \in \mathbb{F}_2^k$.  Let the codeword DM  corresponding  to this FCC $\mathcal{C}_0$ be denoted $H_0$. The $(i,j)^{\text{th}}$ entry of $H_0$ is  $H_0(i,j) = d_H(\mathbf{u}_i, \mathbf{u}_j)+d_H(\mathbf{p}^{\mathcal{C}_0}(\mathbf{u}_i),\mathbf{p}^{\mathcal{C}_0}(\mathbf{u}_j))$, for all $i,j \in \{1,2,\cdots, 2^k\}$. 

    Now, consider another optimal SEFCC $\mathcal{C}_1$ for the given MUBF $f$ with the corresponding codeword DM denoted as $H_1$. The parity assignment $\mathbf{p}^{\mathcal{C}_1}(\mathbf{u}_i)$ for a vector $\mathbf{u}_i$ in the FCC $\mathcal{C}_1$ is obtained by flipping the first bit in the parity assignment $\mathbf{p}_{\mathcal{C}_0}(\mathbf{u}_i)$ of the FCC $\mathcal{C}_0$ and retaining the second parity bit in $\mathbf{p}_{\mathcal{C}_0}(\mathbf{u}_i)$ as it is, i.e., $\mathbf{p}^{\mathcal{C}_1}(\mathbf{u}_i) = [(p^{\mathcal{C}_0}_1(\mathbf{u}_i))^c, \ p^{\mathcal{C}_0}_2(\mathbf{u}_i)], \forall \mathbf{u}_i \in \mathbb{F}_2^k$. Since, going from $\mathcal{C}_0$ to $\mathcal{C}_{1}$, the first parity bit of all the codewords changes to their respective complements, the distance between two codewords $d_H(enc_{\mathcal{C}_1}(\mathbf{u}_i), enc_{\mathcal{C}_1}(\mathbf{u}_j))$ in the FCC $\mathcal{C}_1$ remains the same as the their corresponding distance w.r.t to the FCC $\mathcal{C}_0$. Thus, the two distance matrices $H_0$ and $H_1$ are the same. 

    Similarly, consider optimal SEFCCs $\mathcal{C}_2$ and $\mathcal{C}_3$ with codeword DMs $H_2$ and $H_3$, respectively, with parity assignments 
    $\mathbf{p}^{\mathcal{C}_2}(\mathbf{u}_i) = [p^{\mathcal{C}_0}_1(\mathbf{u}_i), (p^{\mathcal{C}_0}_2(\mathbf{u}_i))^c]$, and $\mathbf{p}^{\mathcal{C}_3}(\mathbf{u}_i) = [(p^{\mathcal{C}_0}_1(\mathbf{u}_i))^c, (p^{\mathcal{C}_0}_2(\mathbf{u}_i))^c]$, respectively, $\forall \mathbf{u}_i \in \mathbb{F}_2^k$. Since, the code $\mathcal{C}_2$ has the second parity bit flipped and the first retained for all the codewords w.r.t the FCC $\mathcal{C}_0$, their corresponding codeword DMs are the same, i.e., $H_0 = H_2$. Similarly, we have $H_0 = H_3$. 
    
    Thus, we have the codeword DMs corresponding to the optimal SEFCCs $\mathcal{C}_0, \mathcal{C}_1,\mathcal{C}_2,$ and $\mathcal{C}_3$, obtained for each of the four possible parity assignments of the all-zero vector $\mathbf{0}_k$, all the same. Since, with each of the four parity assignments for $\mathbf{0}_k$, there are an equal number of possible optimal SEFCCs, and corresponding to each FCC with a fixed parity assignment $[p_1,\ p_2]$ for $\mathbf{0}_k$, there is an optimal SEFCC with the same codeword DM for each of the other three parity assignments for $\mathbf{0}_k$, we conclude that all distinct codeword DMs for optimal SEFCCs for a given MUBF $f$ will arise with a fixed parity assignment for the all-zero vector $\mathbf{0}_k$. 
\end{IEEEproof}

\textbf{Note:} Hereafter, for any given MUBF $f: \mathbb{F}_2^k \rightarrow \mathbb{F}_2$,  we shall only consider optimal SEFCCs with the parity assignment for the all-zero vector as $\mathbf{p}(\mathbf{0}_k)=[0,\ 0]$. The phrase "total number of possible FCCs" shall be used to refer to the total number of FCCs with the above parity assignment for the all-zero vector $\mathbf{0}_k$ and we denote this number by $\hat{\mathcal{N}}(f,t=1))$. 

\begin{lem}
\label{Lem:CompWt-1}
    For a given MUBF $f: \mathbb{F}_2^k \rightarrow \mathbb{F}_2$, if an optimal SEFCC $\mathcal{C}_1$ is obtained from another optimal SEFCC $\mathcal{C}_0$ by complementing all the parity vectors of weight $1$ in $\mathcal{C}_0$, then the codeword DM of $\mathcal{C}_1$ is the same as that of $\mathcal{C}_0$. 
\end{lem}

\begin{IEEEproof}
    Consider the ${i,j}^{\text{th}}$ entry $H_0(i,j)$ in the codeword DM $H_0$ corresponding to the SEFCC $\mathcal{C}_0$. We have $H_0(i,j) = d_H(\mathbf{u}_i, \mathbf{u}_j)+d_H(\mathbf{p}^{\mathcal{C}_0}(\mathbf{u}_i), \mathbf{p}^{\mathcal{C}_0}(\mathbf{u}_j)$ and $H_1(i,j) = d_H(\mathbf{u}_i, \mathbf{u}_j)+d_H(\mathbf{p}^{\mathcal{C}_1}(\mathbf{u}_i), \mathbf{p}^{\mathcal{C}_1}(\mathbf{u}_j)$, for all $i,j \in \{1,2,\cdots, 2^k\}$. 

    \textit{Case 1}: Consider the case where $\mathbf{p}^{\mathcal{C}_0}(\mathbf{u}_i)$ and $\mathbf{p}^{\mathcal{C}_0}(\mathbf{u}_j)$ are both $[0,\ 0]$ or both $[1,\ 1]$. Then $\mathbf{p}^{\mathcal{C}_0}(\mathbf{u}_i)$ = $\mathbf{p}^{\mathcal{C}_1}(\mathbf{u}_i)$ and $\mathbf{p}^{\mathcal{C}_)}(\mathbf{u}_j) = \mathbf{p}^{\mathcal{C}_1}(\mathbf{u}_j)$ by definition of $\mathcal{C}_1$ and hence, $H_0(i,j) = H_1(i,j)$. 

   \textit{Case 2}: Let $w_H(\mathbf{p}^{\mathcal{C}_0}(\mathbf{u}_i)) = 0$ or $2$, and $w_H(\mathbf{p}^{\mathcal{C}_0}(\mathbf{u}_j)) = 1$. Then, $\mathbf{p}^{\mathcal{C}_0}(\mathbf{u}_i) = \mathbf{p}^{\mathcal{C}_1}(\mathbf{u}_i)$ and $\mathbf{p}^{\mathcal{C}_1}(\mathbf{u}_j) = (\mathbf{p}^{\mathcal{C}_0}(\mathbf{u}_i))^c$. However, since the complement of a weight-$1$ vector of length $2$ is also a weight-$1$ vector of length $2$, and the distance from $\mathbf{0}_2$ or from $\mathbf{1}_2$ to both $[0,1]$ and $[1,0]$ is $1$, $H_0(i,j) = H_1(i,j)$. 

   \textit{Case 3}: Let both $w_H(\mathbf{p}^{\mathcal{C}_0}(\mathbf{u}_i)) = 1$, and $w_H(\mathbf{p}^{\mathcal{C}_0}(\mathbf{u}_j)) = 1$. Then, in $\mathcal{C}_1$, both the parity vectors change to their respective complements, keeping the distance between them the same as that in $\mathcal{C}_0$. 
\end{IEEEproof}

\begin{thm}
\label{Thm:DistinctDMs}
     For a given MUBF $f: \mathbb{F}_2^k \rightarrow \mathbb{F}_2$, the number of optimal SEFCCs with distinct codeword DMs is equal to 
    $$\mathcal{N}_{dis}(f,t=1)=\dfrac{\hat{\mathcal{N}}(f,t=1)) + 2^{(2^k - \binom{k}{2} - \binom{k}{1} - \binom{k}{0})}}{2},$$
    where $\hat{\mathcal{N}}(f,t=1)) = \dfrac{\mathcal{N}(f,t=1)}{4}$ is the number of optimal SEFCCs with the parity vector $\mathbf{0}_2$ for the all-zero vector $\mathbf{0}_k$. 
\end{thm}

\begin{IEEEproof}
    We consider all optimal SEFCCs for the given MUBF $f$ with the parity vector $\mathbf{0}_2$ for the all-zero vector $\mathbf{0}_k$. In all these FCCs, the parity vectors for all the $\binom{k}{1}$ weight-$1$ vectors must be $\mathbf{1}_2$ to satisfy the distance requirement.

    Now, divide the SEFCCs into two groups. Group 1 shall contain all FCCs with at least one parity vector of Hamming weight $1$ whereas Group 2 shall contain all FCCs with no weight-$1$ parity vectors. Any FCC in Group 1 has another FCC in Group 1 such that they have the same codeword DM by Lemma \ref{Lem:CompWt-1}. Hence, the number of optimal SEFCCs with distinct codeword DMs is upper bounded by 
$\frac{|\text{Group 1}|}{2} + |\text{Group 2}|$, 
where $|\text{Group i}|$ denotes the number of SEFCCs in Group $i$ for $i = 1,2$.

\begin{figure*}[t]
    \centering
    \captionsetup{justification=centering}
    \includegraphics[width=\textwidth]{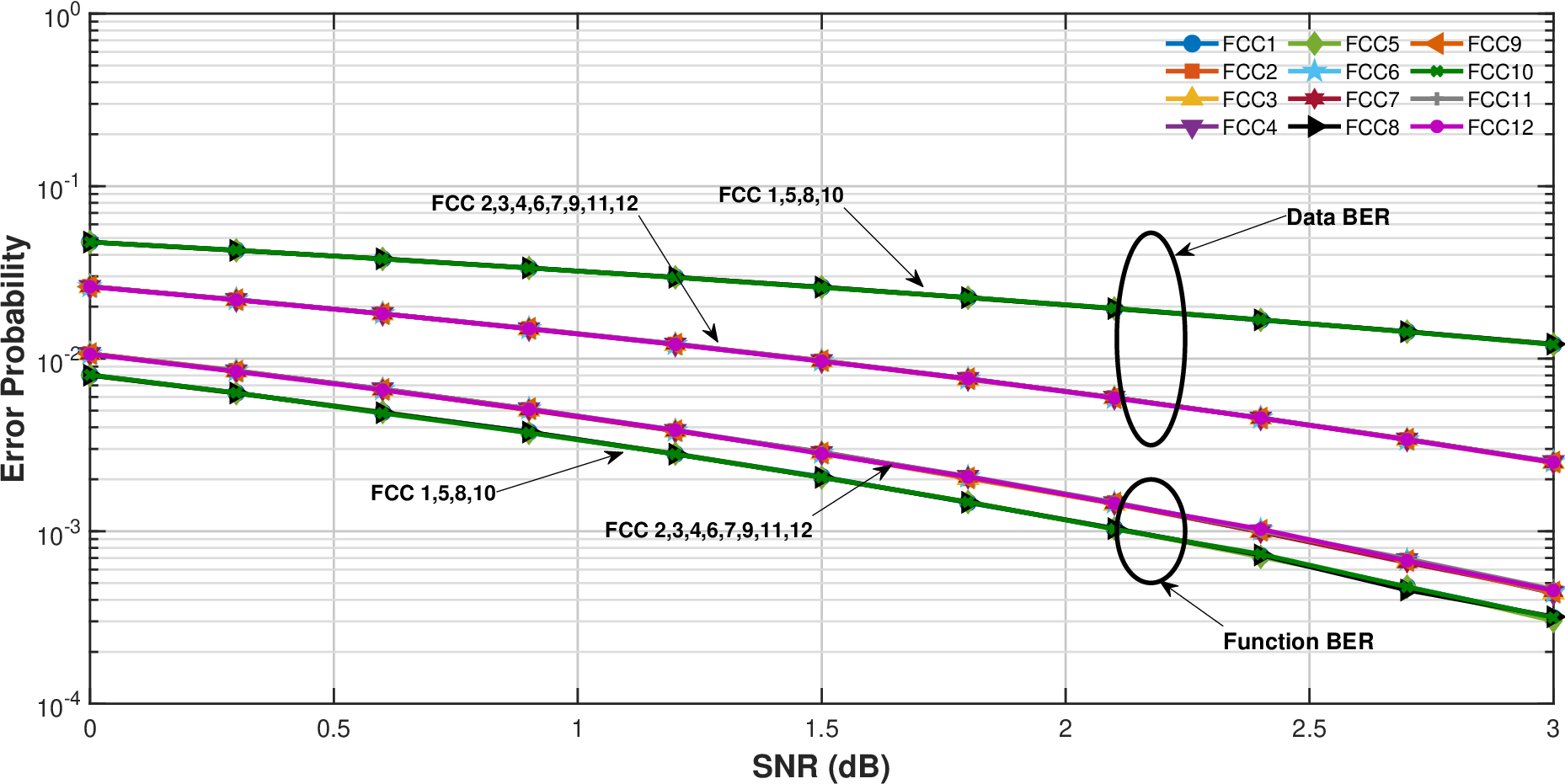}
  \caption{Error performance of FCCs for OR function with $k=2$ under soft-decision decoding.}
    \label{soft_12}
\end{figure*}

\begin{figure*}[t]
    \centering
    \captionsetup{justification=centering}
    \includegraphics[width=\textwidth]{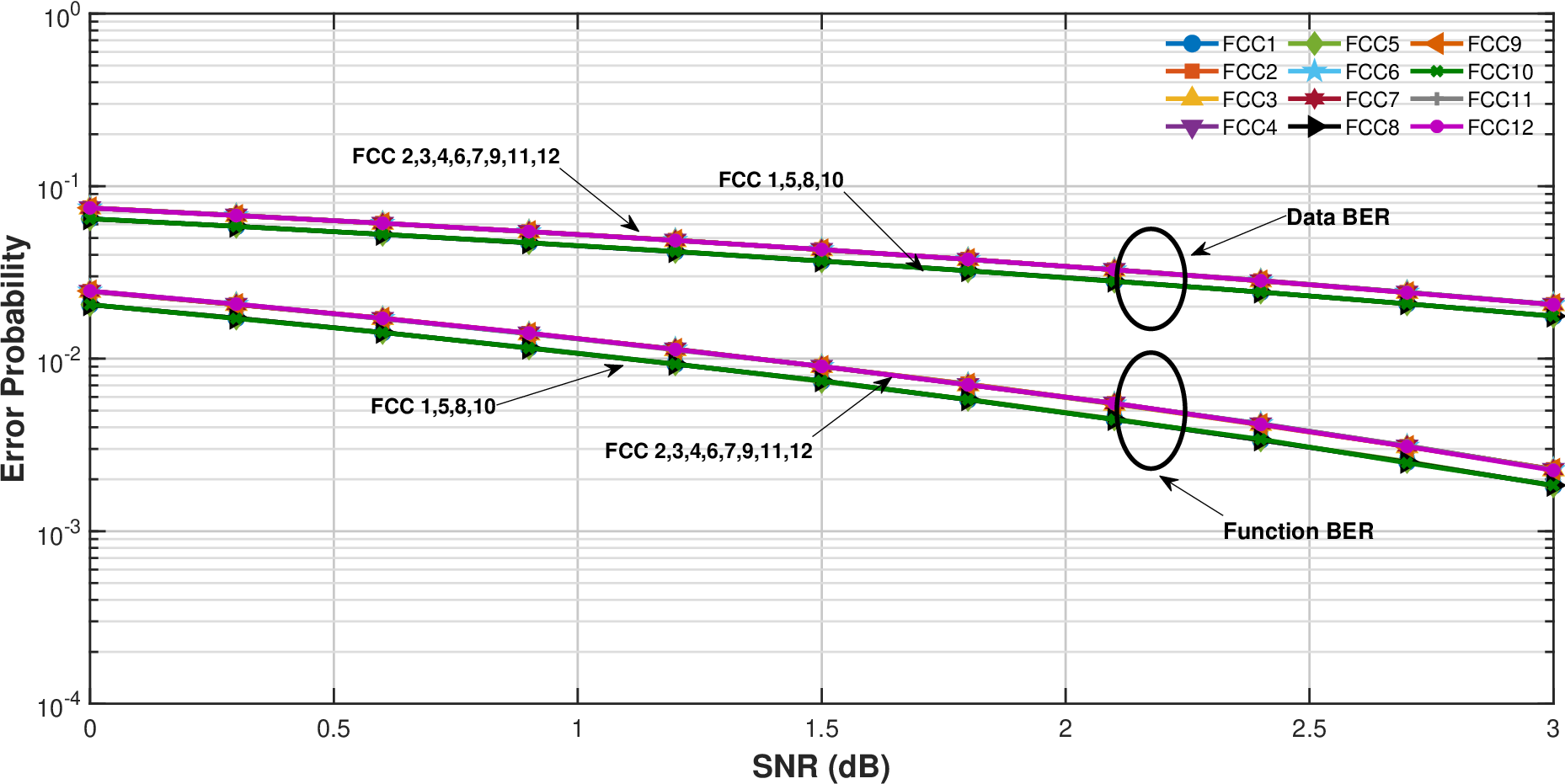}
  \caption{Error performance of FCCs for OR function with $k=2$ under hard-decision decoding.}
    \label{hard_12}
\end{figure*}

    Now we shall prove that this upper bound is the exact number of SEFCCs with distinct codeword DMs. Assume for the sake of contradiction that there exist two distinct SEFCCs $\mathcal{C}_0$ and $\mathcal{C}_1$, both of which have the same codeword DM, but they don't form a pair as described in Lemma \ref{Lem:CompWt-1}. Since $\mathcal{C}_0$ and $\mathcal{C}_1$ are distinct, there must exist at least one vector $u_i \in \mathbb{F}_2^k$ such that its parity vector is different in both the FCCs, i.e., $\mathbf{p}^{\mathcal{C}_0}(\mathbf{u}_i) \neq \mathbf{p}^{\mathcal{C}_1}(\mathbf{u}_i)$.

    \textit{Case 1:}
    If the Hamming weights of these parity vectors are different, i.e., $w_H(\mathbf{p}^{\mathcal{C}_0}(\mathbf{u}_i)) \neq w_H(\mathbf{p}^{\mathcal{C}_1}(\mathbf{u}_i))$, then, they would have different distances to the parity vector $\mathbf{0}_2$ assigned to the all-zero input vector $\mathbf{0}_k$ in both codes $\mathcal{C}_0$ and $\mathcal{C}_1$, which contradicts our assumption that the codeword DMs of $\mathcal{C}_0$ and $\mathcal{C}_1$ are the same. 

    \textit{Case 2:}
    If the Hamming weights of these parity vectors are the same, i.e., $w_H(\mathbf{p}^{\mathcal{C}_0}(\mathbf{u}_i)) = w_H(\mathbf{p}^{\mathcal{C}_1}(\mathbf{u}_i))$, then one of them has to be $[0,\ 1]$ and the other $[1,\ 0]$. 
    
    \textit{Case 2a:}If there does not exist another vector  $\mathbf{u}_j \in \mathbb{F}_2^k$, $\mathbf{u}_j \neq \mathbf{u}_i$ such that its parity vector is the same weight-1 vector in both $\mathcal{C}_0$ and $\mathcal{C}_1$, then $\mathcal{C}_0$ and $\mathcal{C}_1$ form a pair as defined in Lemma \ref{Lem:CompWt-1} contradicting our assumption. 
    
    \textit{Case 2b:} If there exists a vector  $\mathbf{u}_j \in \mathbb{F}_2^k$, $\mathbf{u}_j \neq \mathbf{u}_i$ such that its parity vector is the same weight-1 vector $\mathbf{p}(\mathbf{u}_j)$ in both $\mathcal{C}_0$ and $\mathcal{C}_1$, then the distance of  $\mathbf{p}(\mathbf{u}_j)$ from $\mathbf{p}^{\mathcal{C}_0}(\mathbf{u}_i)$ and $\mathbf{p}^{\mathcal{C}_1}(\mathbf{u}_i)$ are different contradicting our assumption that $\mathcal{C}_0$ and $\mathcal{C}_1$ have the same codeword DMs. 
    Since every possible case results in a contradiction to our assumption, the number of optimal SEFCCs with distinct codeword DMs is $\dfrac{|\text{Group 1}|}{2} + |\text{Group 2})| = \dfrac{\hat{\mathcal{N}}(f,t=1))-|\text{Group 2}|}{2} + |\text{Group 2})|$

   Group 2 contains all those optimal SEFCCs with no weight-1 parity assignment. So, following the proof of Theorem \ref{Thm:NumFCC}, fixing $\mathbf{p}(\mathbf{0}_k)$ as $\mathbf{0}_2$ fixes the parity vector for all weight-$1$ input vectors as $\mathbf{1}_2$, For all weight-$2$ vectors, the only possibility for the parity vector is $\mathbf{1}_2$ when weight-1 vectors are not allowed. Every other vector can have the parity vector as either $\mathbf{0}_2$ or $\mathbf{1}_2$ resulting in $|\text{Group 2}| = 2^{(2^k - \binom{k}{2} - \binom{k}{1} - \binom{k}{0})}$.
\end{IEEEproof}

\begin{thm}
For a MUBF $f: \mathbb{F}_2^k \rightarrow \mathbb{F}_2$, the number of optimal SEFCCs with distinct, incomparable, chain leader codeword DMs is given by $$\mathcal{N}_{CL}(f,t=1) = 2^\alpha3^\beta + 2^\beta(1 + \alpha + \frac{\beta}{2}),$$ where $\alpha = \binom{k}{2}$ and $\beta = 2^k - \binom{k}{2} - \binom{k}{1} - \binom{k}{0}$.

\end{thm}

\begin{table}[!t]
\centering
\scriptsize
\setlength{\tabcolsep}{4pt}
\renewcommand{\arraystretch}{1.1}
\begin{tabular}{cccc|cccc|cccc}
\multicolumn{4}{c}{FCC 1} & \multicolumn{4}{c}{FCC 2} & \multicolumn{4}{c}{FCC 3} \\
$u_1$ & $u_2$ & $p_1$ & $p_2$ & $u_1$ & $u_2$ & $p_1$ & $p_2$ & $u_1$ & $u_2$ & $p_1$ & $p_2$ \\
0 & 0 & 0 & 0 & 0 & 0 & 0 & 0 & 0 & 0 & 0 & 0 \\
0 & 1 & 1 & 1 & 0 & 1 & 1 & 1 & 0 & 1 & 1 & 1 \\
1 & 0 & 1 & 1 & 1 & 0 & 1 & 1 & 1 & 0 & 1 & 1 \\
1 & 1 & 1 & 1 & 1 & 1 & 1 & 0 & 1 & 1 & 0 & 1 \\[3pt]

\multicolumn{4}{c}{FCC 4} & \multicolumn{4}{c}{FCC 5} & \multicolumn{4}{c}{FCC 6} \\
$u_1$ & $u_2$ & $p_1$ & $p_2$ & $u_1$ & $u_2$ & $p_1$ & $p_2$ & $u_1$ & $u_2$ & $p_1$ & $p_2$ \\
0 & 0 & 0 & 1 & 0 & 0 & 0 & 1 & 0 & 0 & 0 & 1 \\
0 & 1 & 1 & 0 & 0 & 1 & 1 & 0 & 0 & 1 & 1 & 0 \\
1 & 0 & 1 & 0 & 1 & 0 & 1 & 0 & 1 & 0 & 1 & 0 \\
1 & 1 & 0 & 0 & 1 & 1 & 1 & 0 & 1 & 1 & 1 & 1 \\[3pt]

\multicolumn{4}{c}{FCC 7} & \multicolumn{4}{c}{FCC 8} & \multicolumn{4}{c}{FCC 9} \\
$u_1$ & $u_2$ & $p_1$ & $p_2$ & $u_1$ & $u_2$ & $p_1$ & $p_2$ & $u_1$ & $u_2$ & $p_1$ & $p_2$ \\
0 & 0 & 1 & 0 & 0 & 0 & 1 & 0 & 0 & 0 & 1 & 0 \\
0 & 1 & 0 & 1 & 0 & 1 & 0 & 1 & 0 & 1 & 0 & 1 \\
1 & 0 & 0 & 1 & 1 & 0 & 0 & 1 & 1 & 0 & 0 & 1 \\
1 & 1 & 0 & 0 & 1 & 1 & 0 & 1 & 1 & 1 & 1 & 1 \\[3pt]

\multicolumn{4}{c}{FCC 10} & \multicolumn{4}{c}{FCC 11} & \multicolumn{4}{c}{FCC 12} \\
$u_1$ & $u_2$ & $p_1$ & $p_2$ & $u_1$ & $u_2$ & $p_1$ & $p_2$ & $u_1$ & $u_2$ & $p_1$ & $p_2$ \\
0 & 0 & 1 & 1 & 0 & 0 & 1 & 1 & 0 & 0 & 1 & 1 \\
0 & 1 & 0 & 0 & 0 & 1 & 0 & 0 & 0 & 1 & 0 & 0 \\
1 & 0 & 0 & 0 & 1 & 0 & 0 & 0 & 1 & 0 & 0 & 0 \\
1 & 1 & 0 & 0 & 1 & 1 & 0 & 1 & 1 & 1 & 1 & 0 \\
\end{tabular}
\caption{All 12 valid FCCs for $k=2, r=2, t=1$.}
\label{tab:all_fccs}
\end{table}

\subsection{Illustrative Example: $k=2$}

To provide intuition on how the distance matrix influences FCC performance, we first consider a simple case with $k=2$, $r=2$, and $t=1$. 
According to Theorem~\ref{Thm:NumFCC}, the total number of optimal SEFCCs for a given $k$ and $t$ can be computed in general. 
Specializing this result to the present parameters, we enumerate all valid FCCs and obtain a total of 12 codes, listed in Table~\ref{tab:all_fccs}, where each block specifies the message symbols $(u_1,u_2)$ and the parity symbols $(p_1,p_2)$. 
This enumeration shows that, even for specific parameters, several structurally distinct FCCs exist, underscoring the importance of code selection in determining error performance.
The corresponding DRM is

\[
\begin{bmatrix}
0 & 2 & 2 & 1\\
2 & 0 & 0 & 0\\
2 & 0 & 0 & 0\\
1 & 0 & 0 & 0
\end{bmatrix},
\quad \sum_{j} d_{1j}=5,\;\; \sum_{i<j} d_{ij}=5.\\[1ex]
\]

Based on their codeword distance matrices, the 12 FCCs fall into two groups, summarized in Table~\ref{tab:dm_groups}, which is consistent with the characterization of distinct distance matrices given in Theorem~\ref{Thm:DistinctDMs}.

\begin{figure*}[t]
    \centering
    \captionsetup{justification=centering}
    \includegraphics[width=\textwidth]{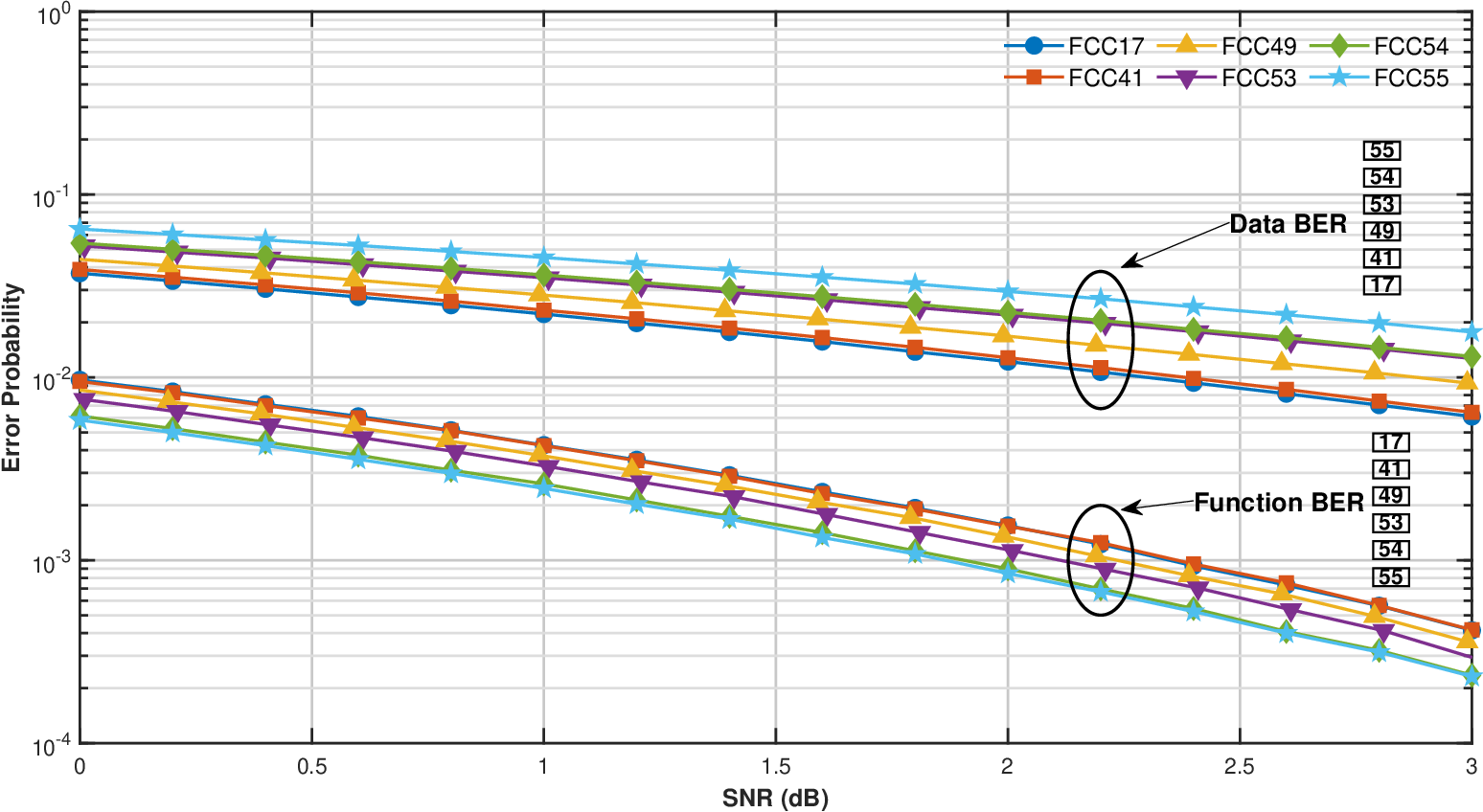}
    \caption{Error performance of FCCs for OR function with $k=3$ for soft decision decoding.}
    \label{fig:k_3soft}
\end{figure*}

Fig.~\ref{soft_12} illustrates the data bit error rate (BER) and function bit error rate (Func BER) of FCCs as a function of the signal-to-noise ratio (SNR) under soft-decision decoding.

It can be observed that FCCs sharing the same codeword distance matrix exhibit nearly identical error performance across the entire SNR range under both soft-decision and hard-decision decoding. As the SNR increases, the error curves of FCCs within the same group converge, demonstrating that the distance distribution fully determines the error performance for each decoding strategy.

\begin{table}[h]
\centering
\caption{Grouping of FCCs by distance matrices for $k=2$, $r=2$, $t=1$.}
\label{tab:dm_groups}
\begin{tabular}{c c}
\toprule
Group 1 (FCCs 2,3,4,6,7,9,11,12) &
Group 2 (FCCs 1,5,8,10) \\
\midrule
$\begin{bmatrix}
0&3&3&3\\
3&0&2&2\\
3&2&0&2\\
3&2&2&0
\end{bmatrix}$ &
$\begin{bmatrix}
0&3&3&4\\
3&0&2&1\\
3&2&0&1\\
4&1&1&0
\end{bmatrix}$ \\
\bottomrule
\end{tabular}
\end{table}
For Group~1, the first row sum is $\sum_j d_{1j} = 9$ and the upper-diagonal sum is $\sum_{i<j} d_{ij} = 15$, 
while for Group~2 these values are $10$ and $14$, respectively. 
As observed in Fig.~\ref{soft_12}, a larger $\sum_j d_{1j}$ (as in Group~2) corresponds to stronger protection against function errors, 
whereas a larger $\sum_{i<j} d_{ij}$ (as in Group~1) leads to improved Data BER performance. 
This explains why FCCs in Group~2 show better resilience in terms of Func BER, 
whereas FCCs in Group~1 perform better in terms of Data BER.

Interestingly, the relationship between the distance matrix and performance depends on the decoding method. 
Under soft-decision decoding, Group~1 (with $\sum_{i<j} d_{ij}=15$) achieves better Data BER than Group~2 (with $\sum_{i<j} d_{ij}=14$), 
which aligns with the intuition that a larger overall distance provides stronger data error protection. 
However, under hard-decision decoding the trend reverses, and Group~2 exhibits lower Data BER, as can be seen in Fig.~\ref{hard_12}. 
This highlights that while the distance matrix captures key structural properties of an FCC, 
the actual performance depends on the decoding strategy: soft-decision decoding fully exploits the Euclidean separation reflected in the sums, 
whereas hard-decision decoding discards this information and may lead to different Data BER and Func BER ordering between groups.\\ Soft-decision decoders use Euclidean distances: larger overall Euclidean separation (upper-diagonal sum) helps, so Group~1 achieves better Data BER.  
Hard-decision first quantizes each received symbol into a bit. The decoder then uses Hamming distance to pick the codeword closest in bit-space. After quantization, the effective error mechanism is single-bit flips (weight-1 patterns dominate). So what matters is how single-bit flips map to decoded codewords and whether they change the function value, not the total Euclidean pairwise sums. Thus, Group~2 can have a smaller upper-diagonal Euclidean sum but a more favorable local Hamming-neighborhood structure (fewer harmful weight-1 error patterns that change decoded bits or Func BER).
\begin{table*}[!t]
\scriptsize
\centering
\begin{tabular}{r l r l}
(1)  & 00000, 00111, 01011, 01101, 10011, 10101, 11001, 11100 & (29) & 00000, 00111, 01011, 01101, 10011, 10111, 11010, 11100 \\
(2)  & 00000, 00111, 01011, 01101, 10011, 10101, 11001, 11101 & (30) & 00000, 00111, 01011, 01101, 10011, 10111, 11010, 11101 \\
(3)  & 00000, 00111, 01011, 01101, 10011, 10101, 11001, 11110 & (31) & 00000, 00111, 01011, 01101, 10011, 10111, 11010, 11110 \\
(4)  & 00000, 00111, 01011, 01101, 10011, 10101, 11001, 11111 & (32) & 00000, 00111, 01011, 01101, 10011, 10111, 11010, 11111 \\
(5)  & 00000, 00111, 01011, 01101, 10011, 10101, 11010, 11100 & (33) & 00000, 00111, 01011, 01101, 10011, 10111, 11011, 11100 \\
(6)  & 00000, 00111, 01011, 01101, 10011, 10101, 11010, 11101 & (34) & 00000, 00111, 01011, 01101, 10011, 10111, 11011, 11101 \\
(7)  & 00000, 00111, 01011, 01101, 10011, 10101, 11010, 11110 & (35) & 00000, 00111, 01011, 01101, 10011, 10111, 11011, 11110 \\
(8)  & 00000, 00111, 01011, 01101, 10011, 10101, 11010, 11111 & (36) & 00000, 00111, 01011, 01101, 10011, 10111, 11011, 11111 \\
(9)  & 00000, 00111, 01011, 01101, 10011, 10101, 11011, 11100 & (37) & 00000, 00111, 01011, 01111, 10011, 10101, 11001, 11100 \\
(10) & 00000, 00111, 01011, 01101, 10011, 10101, 11011, 11101 & (38) & 00000, 00111, 01011, 01111, 10011, 10101, 11001, 11101 \\
(11) & 00000, 00111, 01011, 01101, 10011, 10101, 11011, 11110 & (39) & 00000, 00111, 01011, 01111, 10011, 10101, 11001, 11110 \\
(12) & 00000, 00111, 01011, 01101, 10011, 10101, 11011, 11111 & (40) & 00000, 00111, 01011, 01111, 10011, 10101, 11001, 11111 \\
(13) & 00000, 00111, 01011, 01101, 10011, 10110, 11001, 11100 & (41) & 00000, 00111, 01011, 01111, 10011, 10101, 11010, 11100 \\
(14) & 00000, 00111, 01011, 01101, 10011, 10110, 11001, 11101 & (42) & 00000, 00111, 01011, 01111, 10011, 10101, 11010, 11101 \\
(15) & 00000, 00111, 01011, 01101, 10011, 10110, 11001, 11110 & (43) & 00000, 00111, 01011, 01111, 10011, 10101, 11010, 11110 \\
(16) & 00000, 00111, 01011, 01101, 10011, 10110, 11001, 11111 & (44) & 00000, 00111, 01011, 01111, 10011, 10101, 11010, 11111 \\
(17) & 00000, 00111, 01011, 01101, 10011, 10110, 11010, 11100 & (45) & 00000, 00111, 01011, 01111, 10011, 10101, 11011, 11100 \\
(18) & 00000, 00111, 01011, 01101, 10011, 10110, 11010, 11101 & (46) & 00000, 00111, 01011, 01111, 10011, 10101, 11011, 11101 \\
(19) & 00000, 00111, 01011, 01101, 10011, 10110, 11010, 11110 & (47) & 00000, 00111, 01011, 01111, 10011, 10101, 11011, 11110 \\
(20) & 00000, 00111, 01011, 01101, 10011, 10110, 11010, 11111 & (48) & 00000, 00111, 01011, 01111, 10011, 10101, 11011, 11111 \\
(21) & 00000, 00111, 01011, 01101, 10011, 10110, 11011, 11100 & (49) & 00000, 00111, 01011, 01111, 10011, 10111, 11001, 11100 \\
(22) & 00000, 00111, 01011, 01101, 10011, 10110, 11011, 11101 & (50) & 00000, 00111, 01011, 01111, 10011, 10111, 11001, 11101 \\
(23) & 00000, 00111, 01011, 01101, 10011, 10110, 11011, 11110 & (51) & 00000, 00111, 01011, 01111, 10011, 10111, 11001, 11110 \\
(24) & 00000, 00111, 01011, 01101, 10011, 10110, 11011, 11111 & (52) & 00000, 00111, 01011, 01111, 10011, 10111, 11001, 11111 \\
(25) & 00000, 00111, 01011, 01101, 10011, 10111, 11001, 11100 & (53) & 00000, 00111, 01011, 01111, 10011, 10111, 11011, 11100 \\
(26) & 00000, 00111, 01011, 01101, 10011, 10111, 11001, 11101 & (54) & 00000, 00111, 01011, 01111, 10011, 10111, 11011, 11101 \\
(27) & 00000, 00111, 01011, 01101, 10011, 10111, 11001, 11110 & (55) & 00000, 00111, 01011, 01111, 10011, 10111, 11011, 11111 \\
(28) & 00000, 00111, 01011, 01101, 10011, 10111, 11001, 11111 &     &                                                    \\
\end{tabular}
\caption{Representative FCCs from each of the 55 groups for $k=3$, $r=2$, $t=1$.}
\label{tab:FCC_k3_r2_t1}
\end{table*}
To illustrate these observations for a larger system, we now consider an example with $k=3$.
\begin{example} 
Consider $k = 3$ and the OR function $f(u_1, u_2, u_3) = u_1 \lor u_2 \lor u_3$. Let $u_1, \dots, u_8 \in \mathbb{F}_2^3$ be ordered as follows:
\[
\begin{aligned}
u_1 &= 000, \quad u_2 = 001, \quad u_3 = 010, \quad u_4 = 011, \\
u_5 &= 100, \quad u_6 = 101, \quad u_7 = 110, \quad u_8 = 111.
\end{aligned}
\]
\\
For \( t = 1 \), the Distance Requirement Matrix (DRM) is:

\[
D_f =
\begin{bmatrix}
0 & 2 & 2 & 1 & 2 & 1 & 1 & 0 \\
2 & 0 & 0 & 0 & 0 & 0 & 0 & 0 \\
2 & 0 & 0 & 0 & 0 & 0 & 0 & 0 \\
1 & 0 & 0 & 0 & 0 & 0 & 0 & 0 \\
2 & 0 & 0 & 0 & 0 & 0 & 0 & 0 \\
1 & 0 & 0 & 0 & 0 & 0 & 0 & 0 \\
1 & 0 & 0 & 0 & 0 & 0 & 0 & 0 \\
0 & 0 & 0 & 0 & 0 & 0 & 0 & 0
\end{bmatrix} \\[1ex]
\]

\begin{figure*}[t]
    \centering
    \captionsetup{justification=centering}
    \includegraphics[width=\textwidth]{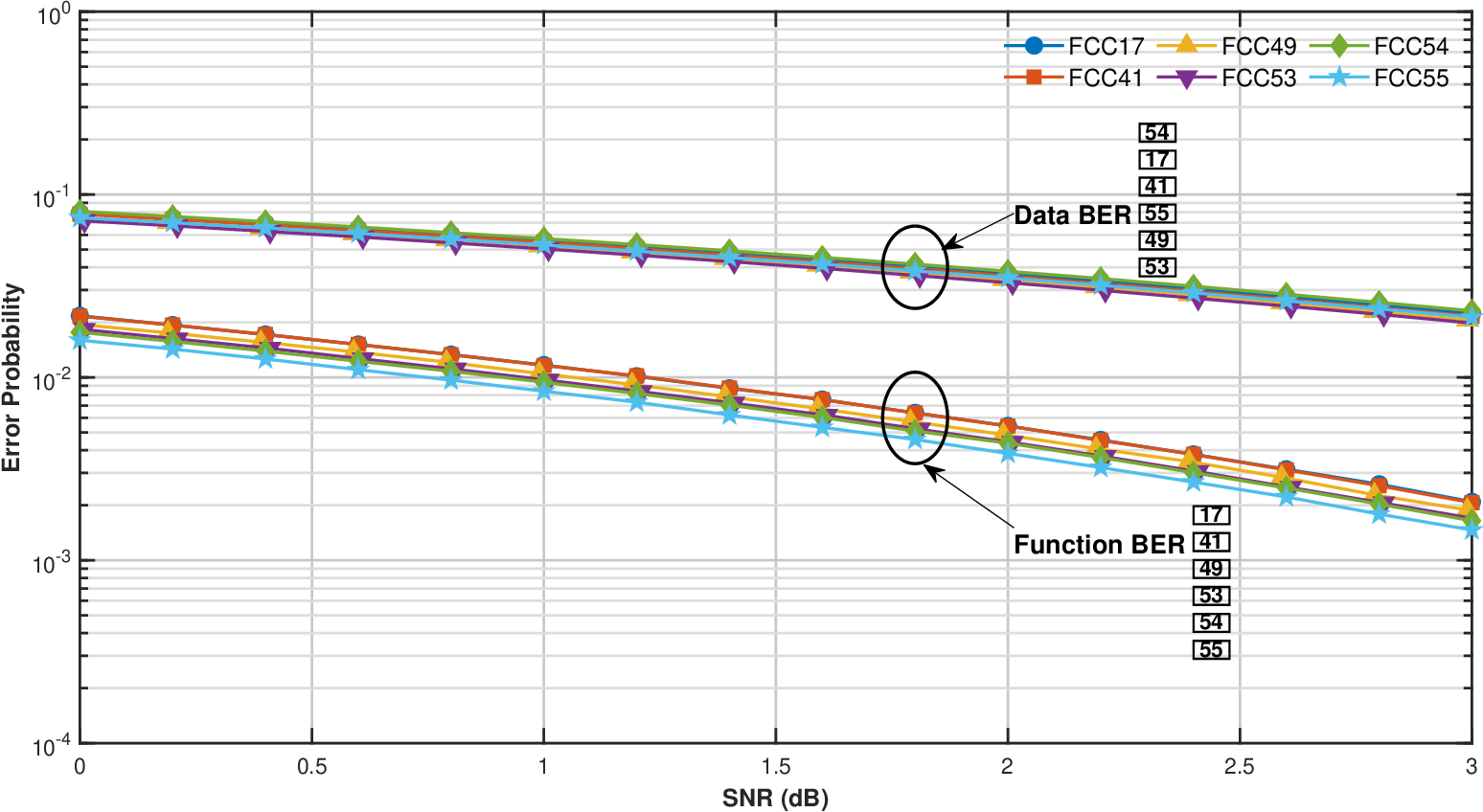}
    \caption{Error performance of FCCs for OR function with $k=3$ for hard decision decoding.}
    \label{fig:k3_hard}
\end{figure*}

Only \( u_1 = 000 \) yields \( f(u_1) = 0 \); for all other inputs, the OR function evaluates to 1. Hence, the distance requirement matrix has non-zero entries only in the first row and column, corresponding to pairs where \( f(u_i) \ne f(u_j) \).

\paragraph{Grouping of FCCs by DM}


\begin{figure*}[t]
    \centering
    \captionsetup{justification=centering}
    \includegraphics[width=\textwidth]{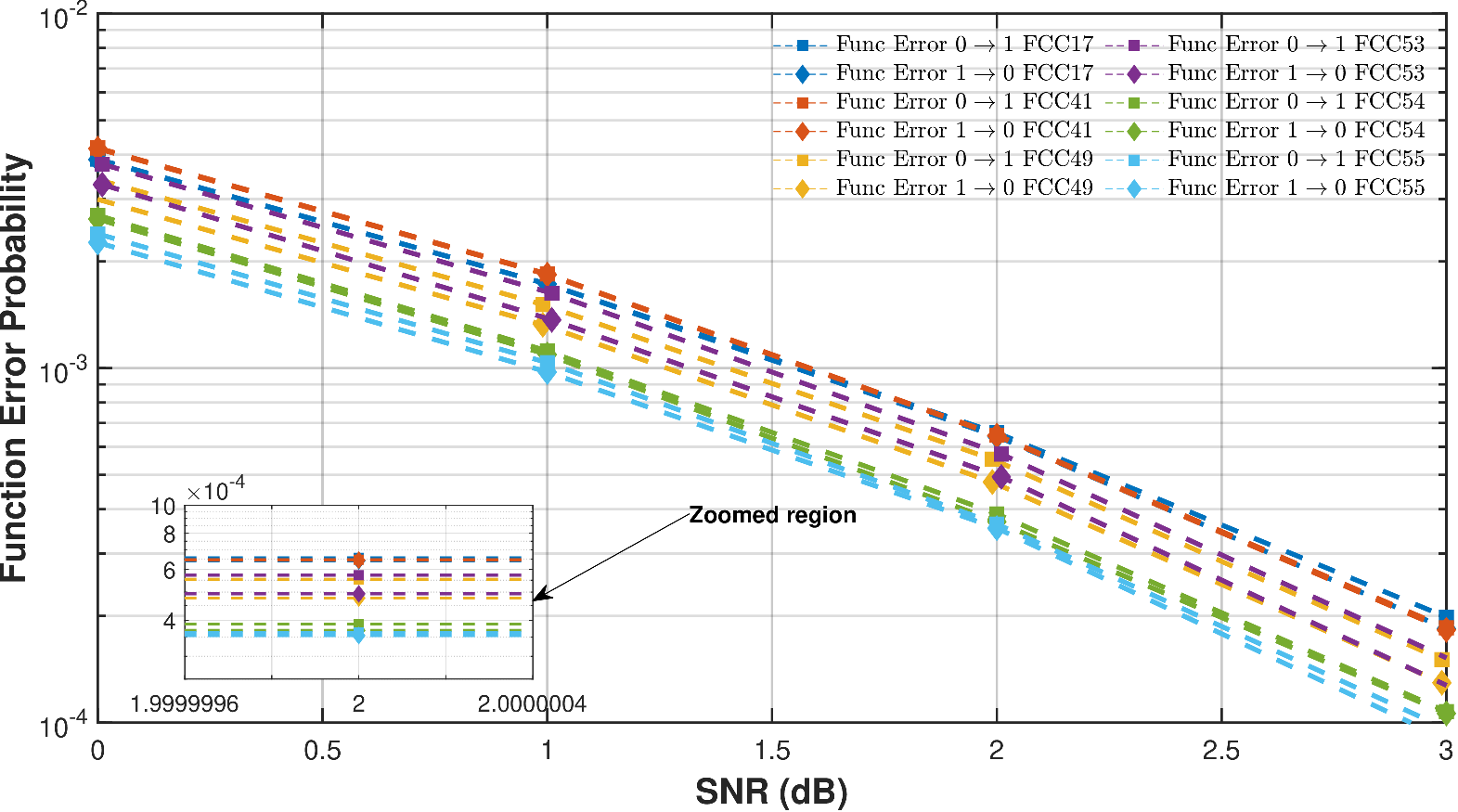}
  \caption{Functional error rate vs.\ SNR for $k=3$, $t=1$ with soft-decision decoding; inset zooms around $\text{SNR}=2$\,dB.}
 \label{fig:k3_01_soft}
\end{figure*}

Although the number of FCCs satisfying the distance constraint grows rapidly with \(k\), as quantified by Theorem~\ref{Thm:NumFCC}, many of these codes are equivalent from a distance-matrix perspective. 
In particular, among the 432 valid FCCs for this setting, only 55 distinct distance matrices arise, in agreement with the grouping predicted by Theorem~\ref{Thm:DistinctDMs}.

For the case of \(k=3\), \(r=2\), and \(t=1\), we construct all valid FCCs satisfying the distance constraint. In total, there are 432 such codes. However, these FCCs fall into only 55 distinct groups, each characterized by a unique distance matrix. Codes within the same group are structurally equivalent with respect to error protection for binary-valued functions, and therefore exhibit comparable decoding performance under both hard-decision and soft-decision decoding. Since it is not feasible to list all 432 codes, Table~\ref{tab:FCC_k3_r2_t1} presents one representative FCC from each of the 55 groups.
\begin{figure*}[!t]
    \centering
    \captionsetup{justification=centering}
    \includegraphics[width=\textwidth]{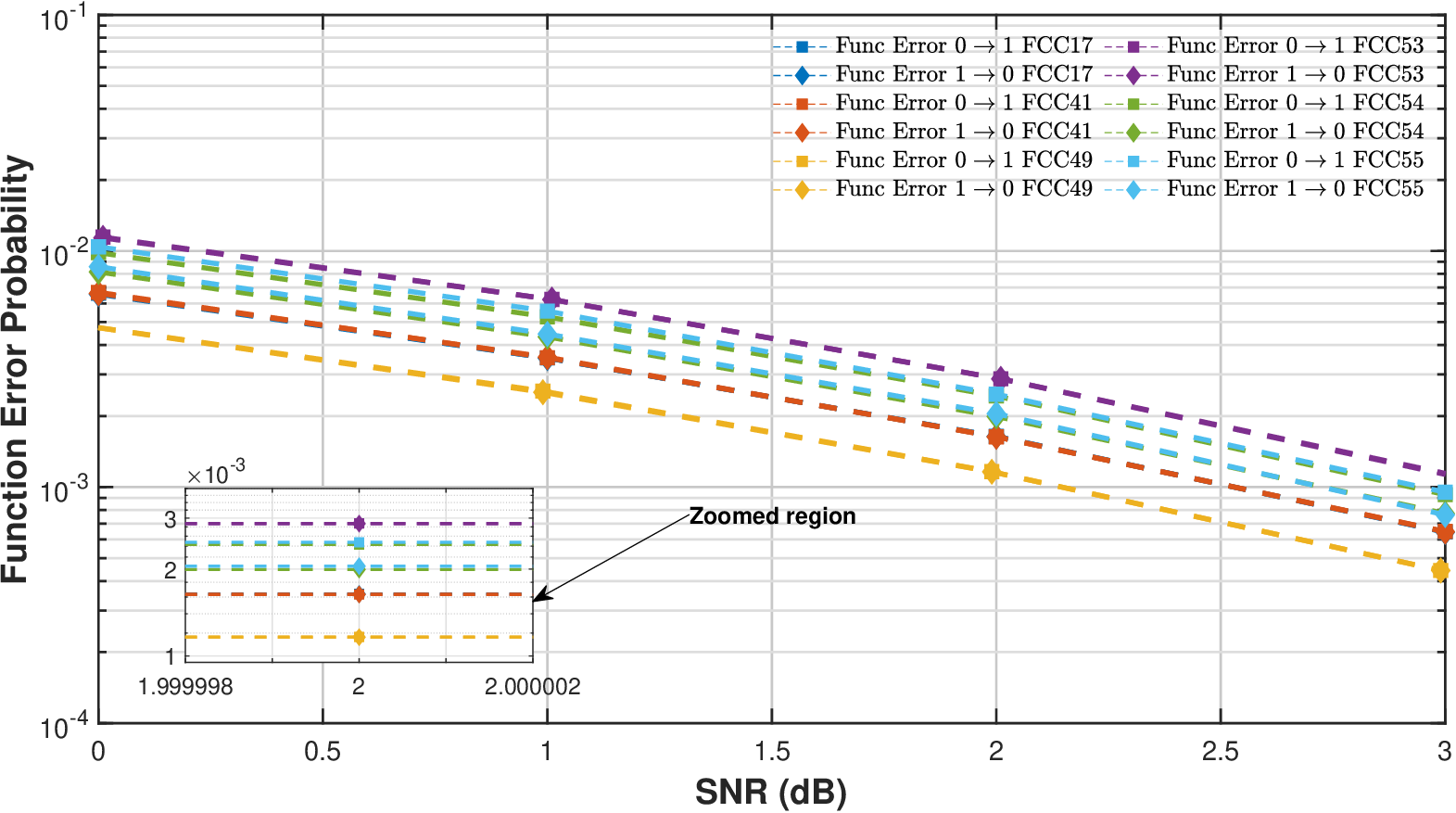}
    \caption{Functional error rate vs.\ SNR for $k=3$, $t=1$ with hard-decision decoding; inset zooms around $\text{SNR}=2$\,dB.}
    \label{fig:k3_01_hard}
\end{figure*}
For example, for the FCC from group~17,
\[
\{00000,\,00111,\,01011,\,01101,\,10011,\,10110,\,11010,\,11100\},
\]
the corresponding distance matrix is

\[
D = \begin{bmatrix}
0 & 3 & 3 & 3 & 3 & 3 & 3 & 3 \\
3 & 0 & 2 & 2 & 2 & 2 & 4 & 4 \\
3 & 2 & 0 & 2 & 2 & 4 & 2 & 4 \\
3 & 2 & 2 & 0 & 4 & 4 & 4 & 2 \\
3 & 2 & 2 & 4 & 0 & 2 & 2 & 4 \\
3 & 2 & 4 & 4 & 2 & 0 & 2 & 2 \\
3 & 4 & 2 & 4 & 2 & 2 & 0 & 2 \\
3 & 4 & 4 & 2 & 4 & 2 & 2 & 0
\end{bmatrix}.
\]
\\

For the case of \(k=3\) and \(t=1\), we selected 6 groups out of the total 55 for our simulations. The selection was based on the distance matrix of each group, specifically considering the first row sum and the upper-diagonal sum. Groups with identical values of both metrics were reduced by retaining only one representative. Among the remaining distinct groups, some exhibited larger first row sums with smaller upper-diagonal sums, while others showed the opposite behavior. Based on this criterion, we finally selected Groups 17, 41, 49, 53, 54, and 55, which are represented by an FCC in Table~\ref{tab:FCC_k3_r2_t1}.

As seen in Fig.~\ref{fig:k_3soft}, which illustrates the soft-decision decoding performance for \(k=3, t=1, r=2\), Groups 17 and 41, having the highest upper-diagonal sums, perform the best in terms of Data BER. On the other hand, Groups 54 and 55, which possess the largest first-row sums, achieve the best performance in terms of Func BER. The overall trend indicates that larger upper-diagonal sums correspond to improved Data BER performance, while larger first-row sums correspond to improved Func BER performance. The ordering of the groups in Fig.~\ref{fig:k_3soft}, from top to bottom, follows this relationship.

For hard-decision decoding (Fig.~\ref{fig:k3_hard}), the Data BER curve from top to bottom corresponds to Groups 54, 17, 41, 55, 49, and 53, with Group~53 achieving the lowest Data BER. Unlike soft-decision decoding, Data BER and Func BER under hard decisions do not correlate monotonically with the first-row or upper-diagonal sums of the distance matrix. This behavior arises from the quantization inherent in hard decisions, which discards reliability information and accentuates differences in the pairwise Hamming-distance distributions among codewords, which are not reflected by the first-row or upper-diagonal sums. Consequently, while the Func BER performance remains largely unchanged between soft- and hard-decision decoding, the Data BER follows the expected correlation with upper-diagonal sums under soft decoding but deviates under hard decoding due to the quantization effect, which changes how single-bit errors impact codeword and function errors.
\\
Apart from the standard Data BER and Func BER plots, we have also plotted the function error probabilities corresponding to the OR function (i.e., the error probabilities for 0-to-1 and 1-to-0 transitions separately) for both soft- and hard-decision decoding in Figs.~\ref{fig:k3_01_soft} and \ref{fig:k3_01_hard}. As can be seen in Fig.~\ref{fig:k3_01_soft}, which corresponds to soft-decision decoding, the ordering of FCCs follows the same trend observed in the general Data BER and Func BER plots: for any given FCC, the probability of a 0-to-1 transition is higher than that of a 1-to-0 transition, consistent with the OR function. In contrast, for hard-decision decoding (Fig.~\ref{fig:k3_01_hard}), this ordering among FCCs is not preserved, although the higher 0-to-1 probability compared to 1-to-0 for each FCC is still observed. This behavior is due to the same effects of hard quantization previously discussed, which alter the effective error mechanism and disrupt the correspondence with the distance-matrix-based ordering.

\end{example} 


\section{Conclusion}
In this work, we studied the design and characterization of optimal function-correcting codes (SEFCCs) for maximally-unbalanced Boolean functions. We presented explicit constructions of such codes, derived closed-form expressions for the total number of optimal SEFCCs, and characterized the number of distinct codeword distance matrices. We also performed simulations for different values of \(k\) and \(t\), analyzing both soft- and hard-decision decoding performance. The simulation results illustrate how the structure of the distance matrix influences Data BER and function error probabilities, providing practical insights into the design and selection of FCCs with reliable error-correcting capabilities.

\section*{Acknowledgment}
This work was supported partly by the Science and Engineering Research Board (SERB) of the Department of Science and Technology (DST), Government of India, through the J.C. Bose National Fellowship awarded to B. Sundar Rajan, by the Ministry of Human Resource Development (MHRD), Government of India, through the Prime Minister’s Research Fellowship (PMRF) awarded to Rajlaxmi Pandey, and by the seed grant provided by the Indian Institute of Technology, Hyderabad to Anjana A. Mahesh.

\end{document}